\documentclass[conference]{IEEEtran}
\IEEEoverridecommandlockouts
\usepackage{amsmath,amssymb,amsfonts}
\usepackage{algorithmic}
\usepackage{graphicx}
\usepackage{textcomp}
\usepackage{xcolor}
\usepackage[style=ieee]{biblatex}
\def\BibTeX{{\rm B\kern-.05em{\sc i\kern-.025em b}\kern-.08em
    T\kern-.1667em\lower.7ex\hbox{E}\kern-.125emX}}
\makeatletter
\newcommand{\linebreakand}{%
  \end{@IEEEauthorhalign}
  \hfill\mbox{}\par
  \mbox{}\hfill\begin{@IEEEauthorhalign}
}
\makeatother

\begin{document}

\title{Continuous Target Speech Extraction: Enhancing Personalized Diarization and Extraction on Complex Recordings\\
}	

\author{\IEEEauthorblockN{1\textsuperscript{st} He Zhao}
\IEEEauthorblockA{\textit{College of ISEE} \\
\textit{Zhejiang University}\\
Zhejiang, China \\
zhao$\_$he@zju.edu.cn}
\and
\IEEEauthorblockN{2\textsuperscript{nd} Hangting Chen}
\IEEEauthorblockA{
\textit{Tencent AI Lab}\\
Shenzhen, China \\
erichtchen@tencent.com}
\and
\IEEEauthorblockN{3\textsuperscript{rd} Jianwei Yu}
\IEEEauthorblockA{
\textit{Tencent AI Lab}\\
Shenzhen, China \\
tomasyu@tencent.com}
\and
\IEEEauthorblockN{4\textsuperscript{th} Yuehai Wang}
\IEEEauthorblockA{\textit{College of ISEE} \\
\textit{Zhejiang University}\\
Zhejiang, China \\
wyuehai@zju.edu.cn}
}

\maketitle

\begin{abstract}
Target speaker extraction (TSE) aims to extract the target speaker's voice from the input mixture. Previous studies have concentrated on high-overlapping scenarios. However, real-world applications usually meet more complex scenarios like variable speaker overlapping and target speaker absence. In this paper, we introduces a framework to perform continuous TSE (C-TSE), comprising a target speaker voice activation detection (TSVAD) and a TSE model. This framework significantly improves TSE performance on similar speakers and enhances personalization, which is lacking in traditional diarization methods. In detail, unlike conventional TSVAD deployed to refine the diarization results, the proposed Attention-target speaker voice activation detection (A-TSVAD) directly generates timestamps of the target speaker. We also explore some different integration methods of A-TSVAD and TSE by comparing the cascaded and parallel methods. 
The framework's effectiveness is assessed using a range of metrics, including diarization and enhancement metrics. Our experiments demonstrate that A-TSVAD outperforms conventional methods in reducing diarization errors. Furthermore, the integration of A-TSVAD and TSE in a sequential cascaded manner further enhances extraction accuracy. Audio demos are available on our demo page\footnote{herbhezhao.github.io/Continuous-Target-Speech-Extraction}.

\end{abstract}


\section{Introduction}

In noisy settings like a cocktail party, various sound sources coexist: overlapping conversations, clinking cutlery, background music, and the reverberation off walls and objects. However, in this acoustic environment, we are able to accurately track the utterance of the loudspeaker of interest. This phenomenon is known as the cocktail party effect. Achieving this intricate task relies on our selective attention mechanism, which enables us to concentrate on the desired speaker's voice amidst the chaos. This subject has captivated researchers for decades. 
Target speaker extraction (TSE) bears a strong resemblance to selective hearing \cite{cherry1953some}. It aims to extract a target speaker's speech signal from multi-talker mixture recordings using clues from a speaker of interest \cite{overview}. Such clues may be spatial location clues of the target speaker, videos of the speaker's lips, or pre-recorded enrollment utterances from which the voice characteristics of the target speaker can be obtained. The audio clue is the most common. Distinct from blind source separation, TSE needs an enrollment from a target speaker, which avoids the permutation problem. 


\begin{figure}[!t]
\centering
\includegraphics[width=8.5cm]{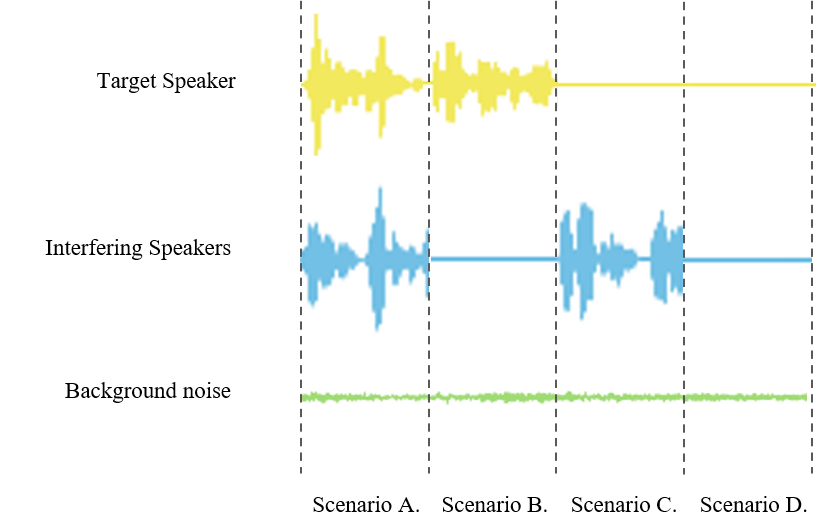}%
\caption{ Input graphical representation of the continuum target speaker extraction task. The active speaker situation can be classified into the following four special scenarios: Scenario A, both the target speaker and the interfering speaker are in active speech; Scenario B, only the target speaker is active and there is no interference from any other speaker; Scenario C, the input mixture contains multiple speakers, but none of them is the target speaker; Scenario D, there is no human voice in the input speech.}
\label{fig0}
\end{figure}

TSE models usually accept a mixture signal and a target-speaker embedding to generate a target-speaker estimation. The target-speaker embedding is calculated using a pre-trained speaker verification model \cite{speakerbeam2017speaker,wang2018voicefilter}. 
Recent researches are mainly based on frequency-domain \cite{li2020atss,wang2020voicefilter,he2022local,ju2023tea} and time-domain \cite{luo2019conv-tasnet,xu2020spex,ge2021multi}. 
Most of the personalized speech enhancement methods focus on highly overlapping and short speech segments, similar to scenarios A and D in Fig.~\ref{fig0}.
In real-world conversations, the rate of speech overlap is approximately 30$\%$ \cite{ccetin2006analysis} and the target speaker may not always be active. However, given the limited number of speakers (\# speaker $\le 3$) in these segments and the occasional absence of the target speaker, current models struggle to produce reasonable outcomes in such scenarios.
For the speaker inactivity problem, it can be solved by enriching the training dataset \cite{borsdorf2021universal} and attaching a speaker judgement module \cite{lin2021sparsely}, but it does not take into account changes in the overlap rate.
According to whether the target speaker and the interfering speakers are active or not, practical application voice scenarios are classified into four conditions, as shown in Fig.~\ref{fig0}.
Typical input audio stream consists of three parts: the target speaker signal, the interfering speakers signal and the background noise. 
Long recordings encompass the aforementioned scenarios and involve a variable number of interfering speakers. Consequently, our research is concentrated on devising strategies for TSE in extended recordings, aiming to enhance their applicability in real-world contexts.

Speaker diarization (SD), which records speaker activity in audio data, is closely related to the task. TSE can be achieved by the active label of target speaker. While diarization effectively identifies non-overlapping speech segments, the performance diminishes in speaker-overlapping regions. To enhance model robustness against complexity, some researchers \cite{taherian2023breaking} have suggested augmenting the TSE loss function with output from a personalized voice activity detector (pVAD). Our research further investigates the synergistic integration of SD and TSE.

In this study, we deal with continuous target speech extraction that is more closely related to everyday life, including more speakers and the complexities of target speaker presence in complex recordings. To address the aforementioned problems, we propose continuous TSE (C-TSE).
C-TSE consists of two sub-networks, one is called personalized band-split RNN (pBSRNN) \cite{yu23b_interspeech}, which is based on BSRNN \cite{luo2023music} with the addition of target speaker registration module for removing interfering speakers signal and background noise. The other is Attention-target speaker voice activation detection (A-TSVAD), which is used to generate high-precision timestamps of the target speaker's active situation. The former focuses on achieving noise reduction and target speaker extraction when the target speaker is present, while the latter removes the interference speech clips directly when the target speaker is inactive. We train the two models separately and test the cascaded and parallel fusion approach between the two.
This framework significantly improves speaker diarization and extraction in such complex recordings. Notably, the C-TSE framework outperformed the TSE model on recordings with a variable number of interference speakers, various overlap ratios and speaker absence. Our contributions are three-fold:

1. \textbf{A-TSVAD} model is introduced with transformer blocks to capture the target speaker's activity, surpassing classic diarization methods.

2. \textbf{C-TSE framework} is proposed by a cascaded fusion of A-TSVAD and pBSRNN results, further promoting the performance on diarization and enhancement metrics.

3. We use multi-disciplinary evaluation indicators to demonstrate the effectiveness of the proposed methodology. Instead of considering only the quality of the extracted speech, we added the ability to handle the absence of the target speaker and the accuracy of the target speaker recognition as a combined measure of model performance. 

The proposed framework is evaluated on a simulated complex dataset, including multi-speakers, variable overlapping ratios, and target speaker absence. The experiment results validate the strength of the proposed framework.

\section{RELATED WORKS}

\subsection{Speaker diarization}

SD resolves the question of "who spoke when", facilitating swift retrieval and localization of specific speaker's fragments. However, clustering-based SD, predicated on the assumption of unique speakers within segments, struggles with overlapping speech. Researchers have proposed end-to-end neural diarization (EEND) \cite{uPIT,EEND-EDA,RSAN} and TSVAD \cite{tsvad}. These studies model the speaker diarization task as a multi-label prediction problem, effectively addressing overlaps by assigning a set of binary labels to each speaker. 
Integrating this task with a selective post-processing module enables initial TSE. Moreover, an additional enhancement module is imperative for mitigating noise and reducing the impact of overlapping interfering speakers.

\subsection{Continuous speech separation}
Continuous Speech Separation (CSS) is the task of generating a set of non-overlapping speech signals from continuous audio that contains multiple discourses with varying degrees of partial overlap. The audio processed in CSS has a lower overall overlap and sparse speaker activation, more accurately reflecting real-world dialogues. The task requires inferring the source embedding from the mixed signal, either through a speaker inference module \cite{shi2020speaker} or clustering \cite{han2021continuous,li2022dual}, and has shown promising results. Our research aims to explore the application of target speaker extraction in these complex scenarios.

\subsection{Integration between target speaker extraction and speaker diarization}

The aims of two key tasks, TSE and SD, are approximated as extracting target speaker speech segments and segmenting speech, respectively.
The EEND-SS \cite{maiti2023eend} directly uses the estimated speech activity of the binarized branches to enhance the quality of the speech separation.
Some studies implemented joint learning of target speech separation and TSVAD \cite{lin2021sparsely}. With the help of TSVAD generation masks, non-target speech is directly muted, which improves the performance of processing sparse overlapping speech. Nevertheless, it does not consider the speaker non-existence case and still faces challenges such as alignment and handling long sentences.
TS-SEP \cite{TS-SEP} has modified TSVAD for the generation of time-frequency masks. Joint TSVAD and TSE trained under a common objective function perform well in speaker separation tasks. But its main goal is to solve the speaker diarization problem rather than target speaker extraction. There are existing studies on the interaction between TSE and SD. The current research does not consider the target speaker extraction task in complex scenarios. We will delve into combining the combination of the two in complex long-form recordings.

\begin{figure*}[!t]
\centering
\includegraphics[width=18cm]{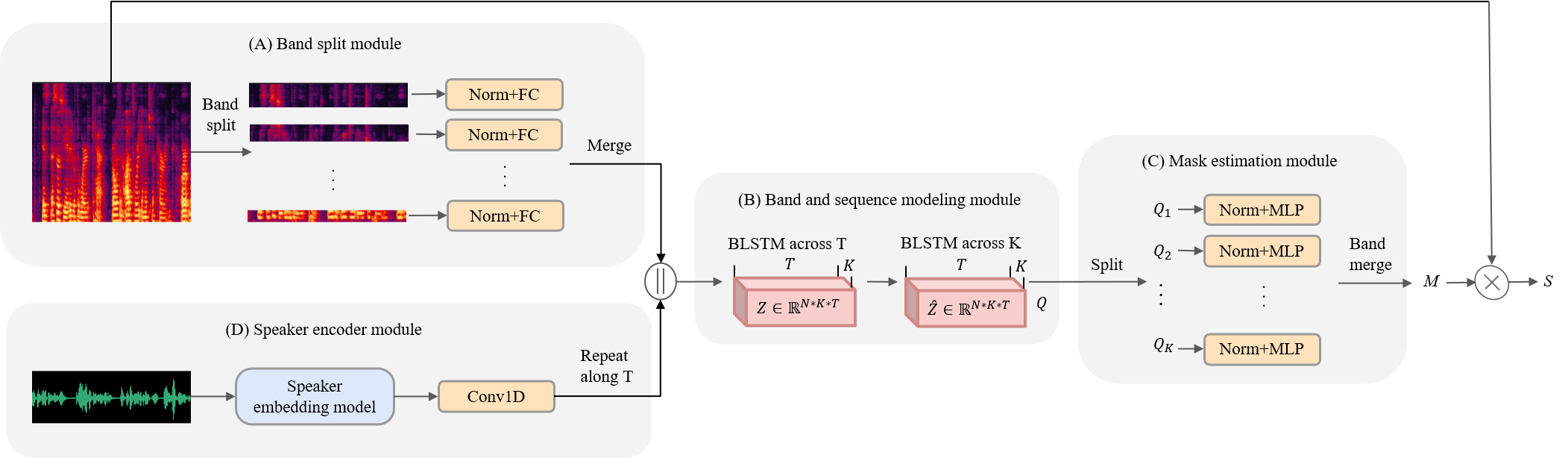}%
\caption{The diagram of the pBSRNN system. (A) The band split module. (B) The band and sequence modeling module. (C) The mask estimation module. (D) The speaker encoder module.}
\label{fig1}
\end{figure*}

\section{Target speaker extraction on complex recordings}

In this paper, we are committed to solving single-channel target speaker extraction. The single-channel observed signal $Y\in \mathbb{R}^T$ is denoted as:
\begin{equation}
\label{deqn_ex1a}
Y=S+I+N
\end{equation}
where $S \in \mathbb{R}^T$ is the target speaker signal, $I \in \mathbb{R}^T$ is the interfering speaker signal, and $N \in \mathbb{R}^T$ is the background noise. $T$ is the number of samples. 

Our framework is composed of three components, including pBSRNN, A-TSVAD and a fusion module. The pBSRNN model aims to extract the signal $S$ from $Y$ with the help of target speaker clues. The pBSRNN first extracts the speaker embedding using a speaker encoder \cite{wespeaker} and then extracts the target speech using BSRNN. We can express the processing
as:
\begin{equation}
\label{deqn_ex1a}
E_S=\text{SpeakerEncoder}(C_S)
\end{equation}
\begin{equation}
\label{deqn_ex1a}
\hat{M}=\text{BSRNN}(Y,E_S)
\end{equation}
where $C_S$ is the target speaker clues, $E_s$ is the clue embeddings and $\hat{M}$ is estimated target-speaker mask. $\text{SpeakerEncoder}(\cdot)$ and $\text{BSRNN}(\cdot)$ respectively represent the target-speaker embedding extraction and the speech enhancement model. 

A-TSVAD predicts the probability $\hat{P}$ of the target-speaker presence with the help of the target speaker clue, which can be denoted as:
\begin{equation}
\label{deqn_ex1a}
\hat{P}=\text{A-TSVAD}(Y,E_S)
\end{equation}
where $ \text{A-TSVAD}(\cdot)$ is the target speaker activity detection model. The $\hat{P}$ is converted to the target speaker active Label by a threshold $\alpha$. 
The process can be represented as:
\begin{equation}
\label{eq5}
\text{Label}=\left\{
\begin{array}{rcl}
0     &      & {\hat{P}      <    \alpha }\\
1     &      & {\hat{P} > \alpha}
\end{array} \right.
\end{equation}


The fusion component fuses the output from pBSRNN and A-TSVAD to generate the final estimation:
\begin{equation}
\label{eq:fusion}
\hat{S}=\text{Fusion}(Y,\hat{M},\hat{P})
\end{equation}
The following sections give a detailed description of pBSRNN, A-TSVAD and the fusion module.

\subsection{pBSRNN}

Fig.~\ref{fig1} is a schematic diagram of the pBSRNN system, which includes band split module, band and sequence modeling module, mask estimation module, and a speaker encoder module.

As shown in Fig.~\ref{fig1} (A), the band split module takes the complex-valued spectrogram generated by the short-time Fourier transform (STFT) as input and segments it into a set of non-overlapping bands based on a predefined bandwidth. This is passed to a layer normalization module and a fully connected (FC) layer to generate a real-valued subband feature. The subband feature is merged to generate the transformed full-band features.
In the speaker encoder module, we employ a ResNet34-based speaker embedding model \cite{wespeaker} that generates speaker embeddings by using a specific 1-D convolutional layers and a $tanh$ activation function and repeated across the time axis. Full band features and speaker embeddings are connected together along the feature dimension.
The band and sequence modeling module is shown in Fig.~\ref{fig1} (B). A bidirectional long short-term memory network (BLSTM) is used to model interleaving at the band and sequence levels of subband features. 
The modeling is first executed through the normalization module, followed by the application of the BLSTM and FC layers. The mask estimation module computes the complex-valued time-frequency (T-F) mask to extract the target source. The T-F mask is generated by first dividing the input features and then by layer normalization and multilayer perceptron (MLP). The masks are merged to obtain a full-band mask 
 $M$ and multiplied with the input complex-valued spectrum to generate the target spectrogram $S$.

 The loss function is defined as the sum of the frequency-domain mean-absolute-error (MAE) loss and a time-domain MAE loss:
 \begin{equation}
\label{eq:fusion}
L=\left \| S-\bar{S}   \right \|_1 +\left \| {\rm iSTFT}\left|  ( S \right ) -{\rm iSTFT}\left ( \bar{S_{r}} \right )   \right \|_1   
\end{equation}
where $\bar{S} \in \mathbb{C}^{F\times T}$ denotes the complex-valued spectrogram of the clean target, and iSTFT denotes the inverse STFT operator.

\subsection{A-TSVAD}

\begin{figure}[!t]
\centering
\includegraphics[width=8.5cm]{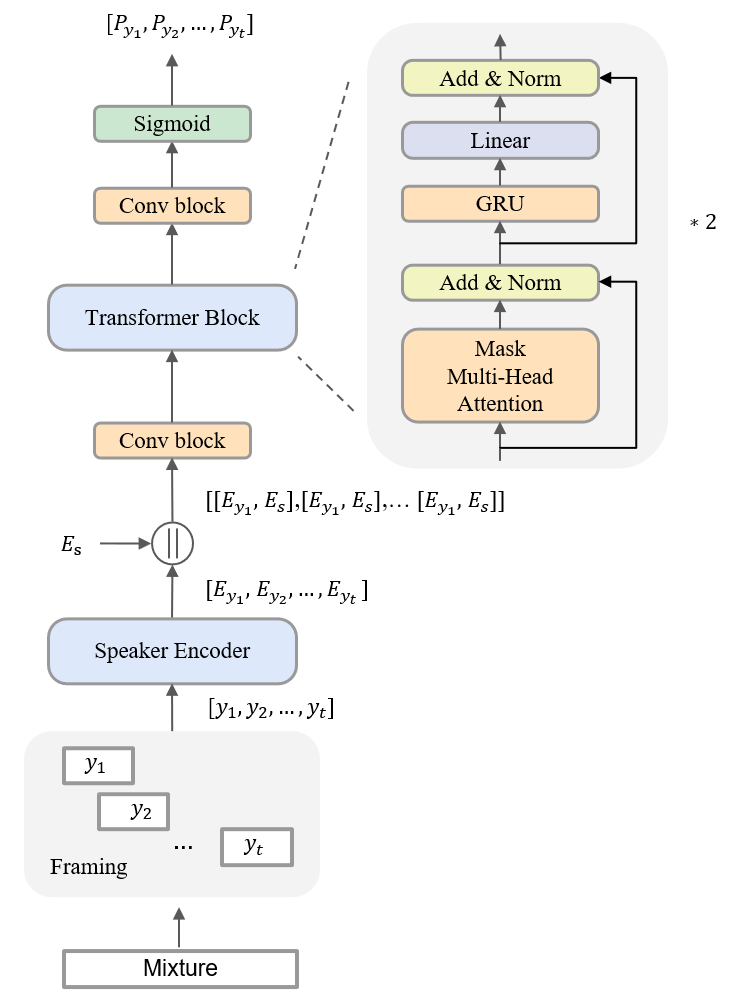}%
\caption{A-TSVAD network}
\label{fig2}
\end{figure}

As shown in Fig.~\ref{fig2}, the A-TSVAD system consists of speaker encoder, convolution blocks and transformer blocks. Firstly, the mixture is framed by a framing block. The frame signals are converted to $[E_{y_1},E_{y_2},...E_{y_t}]$ using speaker encoder. Concatenate it with the target speaker embedding $E_S$ as input.

The convolution blocks consist mainly of one-dimensional convolution and PReLU activation functions. They achieve initial feature integration of the input signal and target-speaker clue. We use 2 stacked transformer blocks. Each block consists of a multi-head attention and a feed-forward layer. The multi-head attention mechanism facilitates the model's capability to concurrently focus on diverse representational subspaces across various positions within the sequence.
The multi-head self-attention is specifically expressed as:
\begin{equation}
\label{deqn_ex1a}
{\rm Attention}(Q,K,V)={\rm Softmax}(\frac{QK^T}{\sqrt{d_k}})V
\end{equation}
\begin{equation}
\label{deqn_ex1a}
\begin{split}
{\rm MultiHead}(Q,K,V)={\rm Concat}({\rm head}_1,\cdots,{\rm head}_h)W^o,\\
{\rm where}\:{\rm head}_i={\rm Attention}(QW_i^Q,KW_i^K,VW_i^V)
\end{split}
\end{equation}
where $Q$, $K$ and $V$ are obtained by passing the input through three linear layers. $d_k$ is the dimension of input, and the projections are parameter matrices $W_i^Q$, $W_i^V$, $W_i^V$ and $W^O$.
The feed-forward module consists of a Gated Recurrent Unit (GRU) and a Linear layer. We employ a residual connection around each of the two sub-layers, followed by layer normalization. 
To enhance stability and facilitate training, a residual connection is incorporated around each sub-layer, succeeded by layer normalization.

The loss function is Binary Cross Entropy (BCE) loss, which measures the distance between the estimated activity and the underlying factual activity of all speakers.
To fill in the small number of silent frames during the active period of the target speaker, we do post-processing smoothing activity estimation on the output probabilities. The post-processing is mainly two operations, dilation and erosion 
 \cite{TS-SEP}:
 \begin{equation}
\label{deqn_ex1a}
\hat{\rm Label} ={\rm Erosion}({\rm Dilation}(\delta(\hat{P}\ge \alpha )))
\end{equation}
where $\hat{\rm Label}$ is a post-smoothing activity estimate, $\alpha$ is the threshold and $\delta(\cdot)$ is shown in Eq.~\ref{eq5}.

\subsection{C-TSE}

\begin{figure}[htb]
\vspace{-1em}
\begin{minipage}[b]{1.0\linewidth}
  \centering
  \centerline{\includegraphics[width=8cm]{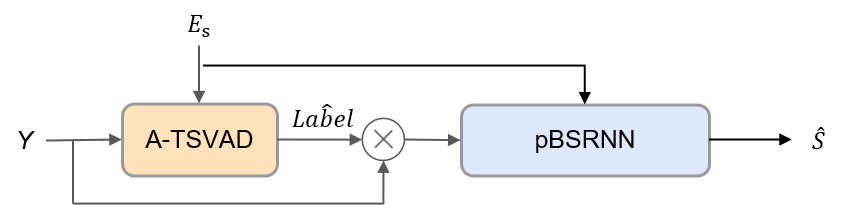}}
  \centerline{(a) Cascade approach 1}\medskip
\end{minipage}
\begin{minipage}[b]{1.0\linewidth}
  \centering
  \centerline{\includegraphics[width=8cm]{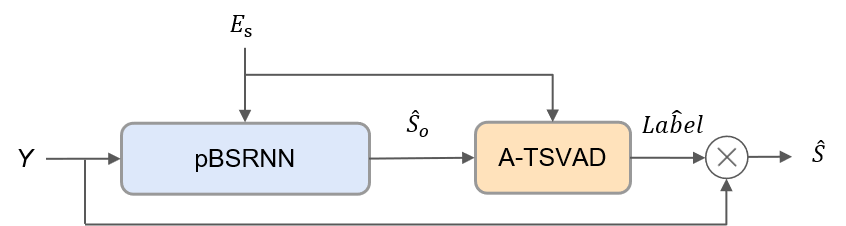}}
  \centerline{(b) Cascade approach 2}\medskip
\end{minipage}
\begin{minipage}[b]{1.0\linewidth}
  \centering
  \centerline{\includegraphics[width=8cm]{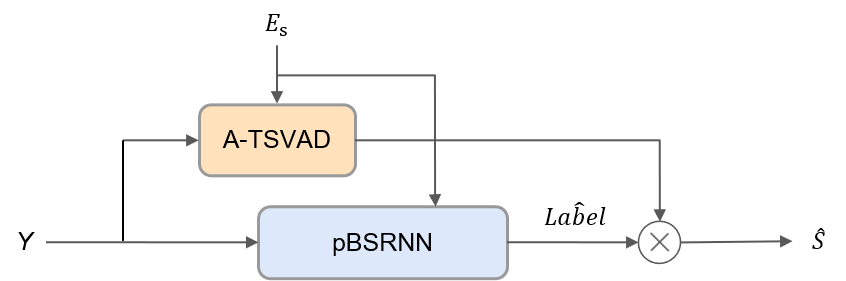}}
  \centerline{(c) Parallel approach}\medskip
\end{minipage}
\caption{Schematic diagram of module integration.}
\label{fig3}
\end{figure}

Fig.~\ref{fig3} illustrates our three proposed approaches for integrating A-TSVAD and pBSRNN, where (a) and (b) are cascaded approaches and (c) is the schema of the parallel approach. Splicing the components in different orders brings specific advantages and limitations. 

In Cascade approach 1, the primary step involves the generation of active labels for the target speaker via A-TSVAD. Subsequently, target speaker active segments are filtered based on these labels, serving as the input to pBSRNN. Optimal performance has been observed when employing a lower A-TSVAD threshold. This minimizes the risk of erroneously filtering out segments where the target speaker is absent.
For Cascade approach 2, the initial extraction of the target speaker is facilitated using pBSRNN. The output from pBSRNN can be characterized as overlap-free speech. While A-TSVAD exhibits enhanced performance in non-overlapping segments, the extraction procedure introduces artifacts, subsequently diminishing A-TSVAD's efficacy.
In the parallel strategy, both models are simultaneously fed with the mixture as input, and their outputs are multiplicatively combined. The interplay between the two models is minimal, leading to a reduced reliance on auxiliary information.


\section{Experimental setup}

\subsection{Dataset}

We stitched together the Librispeech \cite{librispeech} corpus to generate training and test sets. All mixtures were single-channel and sampled at 16 kHz.
As training data, we only performed a 1-minute training session with 2 to 5 randomly selected speakers and an overlap rate of 0-40$\%$. Mute for no more than 3 seconds when there is no overlap between the two speakers. To reduce the effect of channel mismatch, the enrollment speech of the target speaker was convolved with the same room impulse response (RIR) filters \cite{Reddy2020ICASSP2D} as the mixture. Once the splicing is complete, a speaker is randomly selected from it as the target speaker. The order and duration of appearance of the target speakers were randomized. Noise with a randomly selected SNR between 0 and 20 dB was added to the mixture, and the noise is from the DNS challenge \cite{DNS2022}. The training dataset is dynamically generated during training, increasing the number and diversity of samples and improving the robustness of the dataset.

The test set was categorized into six configurations based on their overlap rates: 0L, 0S, OV10, OV20, OV30 and OV40. The configurations 0L and 0S correspond to inter-speech silence durations of 2.9-3 seconds and 0.1-0.5 seconds, respectively. The configuration labeled OV10 indicates a 10$\%$ overlap rate, and similarly, the subsequent configurations (OV20, OV30, OV40) represent increasing overlap rates. Each configuration consists of a 120-second audio mixture involving 2-5 speakers.
\subsection{Model setup}
For the speaker encoder, acoustic features are 80-dimensional log Mel-filterbank energies extracted on frames with a width of 25ms and a step size of 10ms, generating 256-dimensional speaker embeddings.
To add to this, we pre-segment the mixture using a window length of 1.5s and a move of 0.25s before extracting the mixture embedding.

For the pBSRNN, the window and hop size of the Fourier transform are 64ms and 16ms, and the Hanning window is used as the analysis window. The rest of the hyper-parameters were the same as those used in \cite{DBLP:conf/icassp/YuCLGLW23}. Since the pBSRNN training needs to be based on blended speech learning with high overlap, it's divided into two phases of training. The first phase of training is based on mixtures with full overlap and active target speakers to equip it with the ability to extract the target speakers. In the second phase of training, we made the following changes to the training dataset: reducing the overlap of the mixtures, shortening the activity time of the target speakers, and increasing the length of the blended speech. It improves the extraction performance of pBSRNN in complex scenes.

Adam serves as an optimizer with an initial learning rate of $10^{-3}$. The learning rate decays exponentially with a decay floor of 0.99. 
For the A-TSVAD, RAdam serves as the optimizer with an initial learning rate of $10^{-4}$. The learning rate decreased by 0.5 per 10k steps.

\subsection{Compared methods}
We choose as baseline two speaker diarization models, the Bayesian HMM model for clustering x-vector (VBx) \cite{VBx} and Pyannote \cite{pyannote}.
VBx does not require specific model adaptation to any dataset. Pyannote is an open-source toolkit for speaker diaries. Both of them achieve popular off-the-shelf tools for SD.
In addition to this, we trained TSVAD \cite{tsvad} from scratch on the generated data as another baseline to compare the performance of the proposed A-TSVAD. The value of $\alpha$ is 0.025.

The pBSRNN system achieves competitive results in Deep Noise Suppression (DNS) among the various top systems on the DNS2023 challenge \cite{dubey2023icassp}. To analyze the performance of the current target speaker extraction model, we retrain pBSRNN by adding target speaker registered speech as auxiliary information to BSRNN.

\subsection{Performance measurement}
Speaker segmentation accuracy is measured by diarization error rate (DER)
and Jaccard error rate (JER). For both metrics, smaller numbers indicate better performance. 
The DER and JER are as follows:
\begin{equation}
\label{deqn_ex1a}
\text{DER}=\frac{\text{FA}+\text{Miss}+\text{Confusion}}{\text{Total}} 
\end{equation}

\begin{equation}
\label{deqn_ex1a}
\text{JER}=\frac{1}{N}\sum_{i}^{N_\text{ref}}  \frac{\text{FA}_i+\text{Miss}_i}{\text{TOTAL}_i}
\end{equation}
The interference leakage in the absence of the speaker is measured by the INT \cite{taherian2022int}, which calculates the energy difference between the input signal and the remaining signal, which is defined as:
\begin{equation}
\label{deqn_ex1a}
\Delta N=10\log{\left | Y \right |^{2}  } -10\log{\left | \hat{S}  \right |^{2}  }
\end{equation}
where $Y$ is the noisy signal.

Evaluation Metrics Scale-Invariant Signal-to-Noise ratio (SI-SNR) \cite{le2019sisnr} an objective measure of separation accuracy. 
It is formulated as:
\begin{equation}
\label{deqn_ex1a}
\text{SI-SNR}:=10\log_{10}
{\frac{\left \| \frac{\left \langle \hat{s},s  \right \rangle s}{\left \| s \right \|^{2}  }\right \| }
{\left \|\hat{s}- \frac{\left \langle \hat{s},s  \right \rangle s}{\left \| s \right \|^{2}  }\right \| }} 
\end{equation}
where $s,\hat{s} \in \mathbb{R}^T$ are target spealer's clean signals and ertimated signals. $T$ is the audio length.
Perceptual evaluation of speech quality (PESQ)
are used to measure the quality of the extracted speech. For the SI-SNR and PESQ metrics, higher numbers indicate better performance. 

\section{Results and discussion}

\subsection{Baselines and reference results}
The performance of the off-the-shelf models are shown in Table~\ref{table1}. We use the timestamps generated from the speaker diarization to extract the target speaker segments in the mixture.
Owing to the intricacies introduced by overlapping speech, Pyannote, VBx, and TSVAD have inferior DER and JER. The combined metrics of A-TSVAD outperform those of the other two models, and the results show that the currently proposed A-TSVAD outperforms the current clustering-based speaker model and can be useful in providing relatively accurate timestamps.
Furthermore, their SI-SNR and PESQ metrics deteriorate to levels inferior to the baseline mixture. 
Therefore, relying only on the current SD module is not suitable for this case, and it fails to reduce the effects of noise and interfering speakers. 
The SI-SNR and PESQ of pBSRNN can reach 11.1dB and 3.04 with good separation accuracy. However, it still does not have better differentiation accuracy for target speaker segments. 

To summarize, the task of extracting speech from long recorded speech proves challenging when utilizing the off-the-shelf models.

\begin{center} 
\begin{table}[h!]
\vspace{-2em}
\renewcommand\arraystretch{1.5}
\caption{Performance comparison of the off-the-shelf and trained models}
\centering
\resizebox{0.48\textwidth}{!}{
 \begin{tabular} {c c | c c c c c}
 \hline
\multicolumn{2}{c|}{Method} & \multicolumn{5}{c}{Metric}      \\
Timestamp   & Enhancement  & DER$_{\downarrow}$ & JER$_{\downarrow}$ & INT$_{\uparrow}$  & SI-SNR$_{\uparrow}$ & PESQ$_{\uparrow}$ \\
 \hline
 \hline
 N/A & N/A & 256.3 & 62.9 & 0.0  & 6.5 & 2.32\\
 \hline
Pyannote & N/A & 57.4 & 44.5 & 42.0 & 0.5 & 1.33\\
VBx & N/A & 58.7 & 28.1 & 35.3  & 5.1 & 1.90\\
TSVAD & N/A & 43.1 & 34.0 & 26.9 & 3.0 & 1.64\\
 \hline
N/A & pBSRNN & 39.9 & 16.7 & 36.6  & \textbf{11.1} & \textbf{3.04}\\
\hline
\multicolumn{2}{c|}{Cascade approach 1} & \textbf{26.5} & \textbf{16.2} & \textbf{41.3} & 10.4 & 2.91 \\
\multicolumn{2}{c|}{Cascade approach 2} & 84.8 & 44.8 & 25.7 & 2.9 & 2.10\\
\multicolumn{2}{c|}{Parallel approach} & 32.0 & 19.3 & 40.8 & 9.7 & 2.82 \\
\hline
\label{table1}
\end{tabular}}
\vspace{-2em}
\end{table}
\end{center} 

\subsection{Performance of C-TSE}

In our study, we compare the results of C-TSE with other cascade methods. According to the results in Table~\ref{table1}, Cascade approach 1 outperforms other methods in all metrics. It achieves optimal DER and JER and can maintain high INT, SISNR, and PESQ.
The reason for the poorer performance than Cascade approach 1 in the Parallel approach is that the information prior to the two models is not fully utilised.
Cascade approach 1 can combine the advantages of both to enhance the performance of speaker logging and separation in complex recording situations. Not only does it achieve higher audio extraction quality, but also improves the accuracy of recognition of target speaker segments and robustness to the absence of target speakers. 
In particular, Cascade approach 2 introduces a performance degradation using the pBSRNN output estimated by TSVAD. During its cascading process, A-TSVAD needs to be further co-trained with pBSRNN. We will continue to optimise this method in future work. 

\subsection{Performance of TSVAD}

We perform ablation experiments to demonstrate the effectiveness of the improved TSVAD and to analyse the impact of using different window lengths for framing on its overall performance.

\begin{center} 
\begin{table}[h!]
\vspace{-2em}
\renewcommand\arraystretch{1.5}
\caption{Ablation study of TSVAD}
\centering
 \resizebox{0.35\textwidth}{!}{
 \begin{tabular} {c c | c c c}
 \hline
\multicolumn{2}{c|}{TSVAD} & \multicolumn{3}{c}{Metric}      \\
Win   & Arch  & DER$_{\downarrow}$ & JER$_{\downarrow}$ & INT$_{\uparrow}$ \\
 \hline
 \hline
1.50s & BLSTM & 49.3 & \textbf{32.6} & 23.7 \\
\hline
1.50s & Transformer &  \textbf{43.1} & 34.0 & \textbf{26.9} \\
0.75s & Transformer &  74.5 & 74.5 & 29.8 \\
3.00s & Transformer & 55.1 & 52.4 & 28.6 \\
\hline
\label{tableII}
\end{tabular}}
\vspace{-2em}
\end{table}
\end{center} 

\begin{figure}[!t]
\centering
\includegraphics[width=8.5cm]{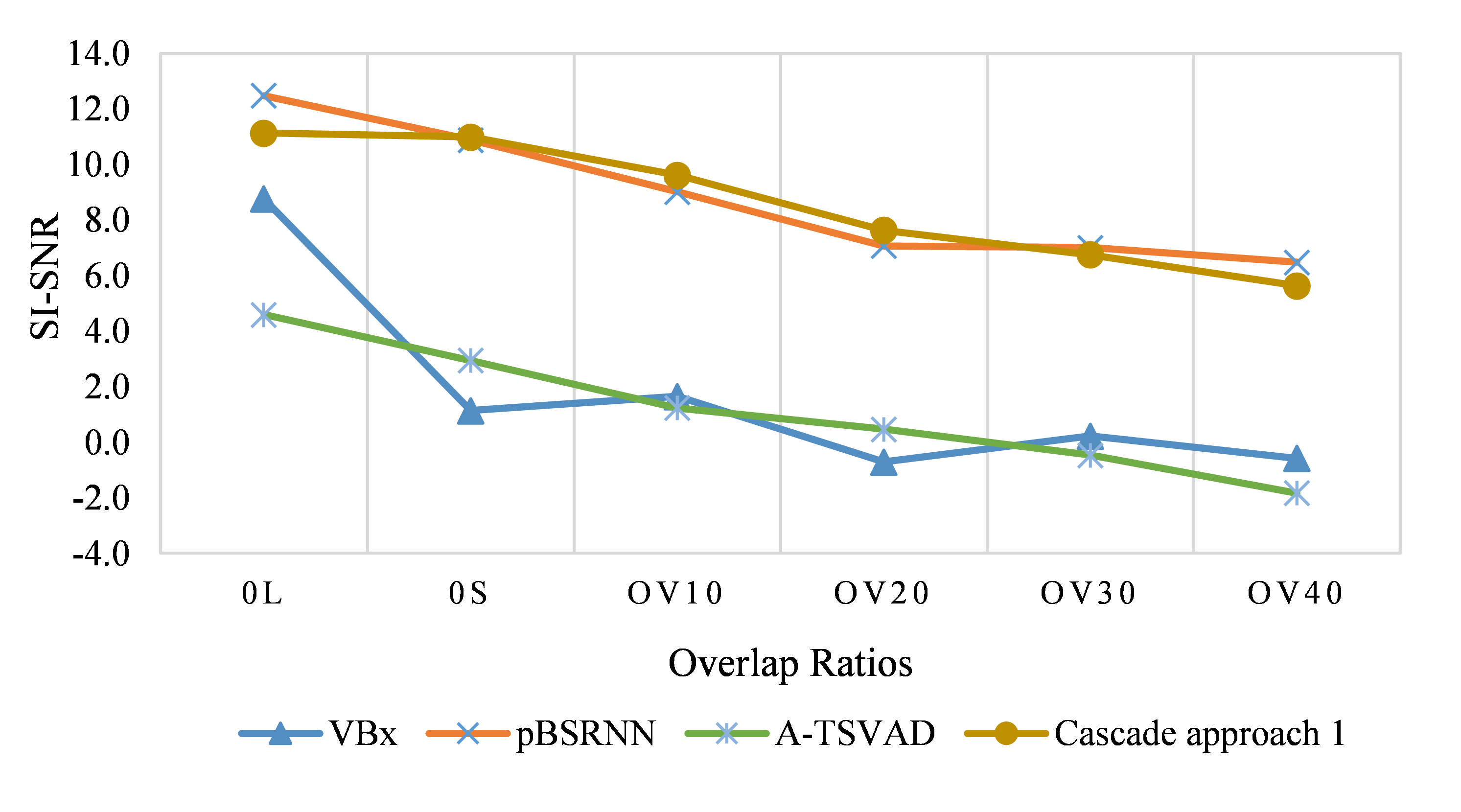}%
\caption{SI-SNR performance of models at different overlap rates}
\label{fig5-1}
\end{figure}

\begin{figure}[!t]
\centering
\includegraphics[width=8.5cm]{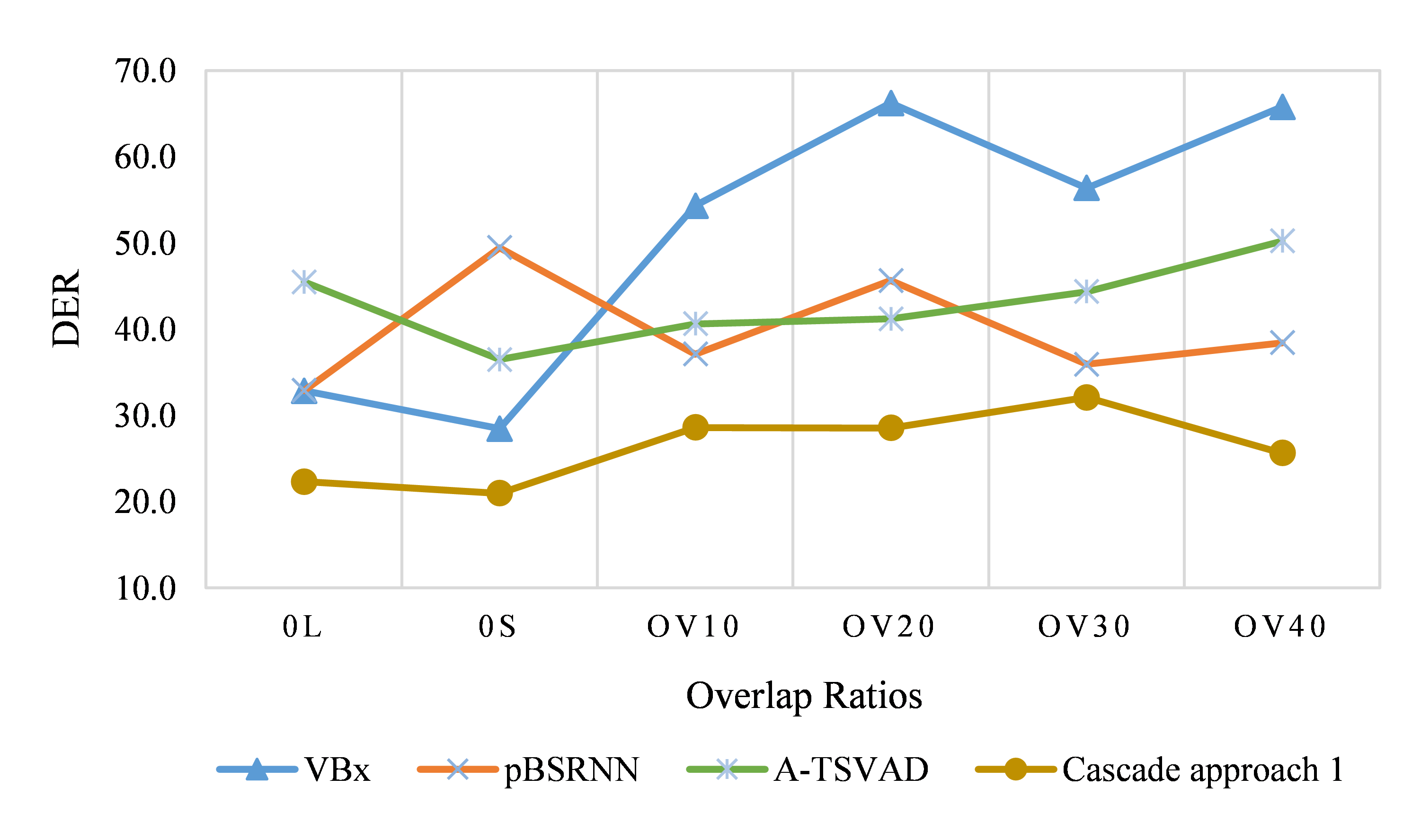}%
\caption{DER performance of models at different overlap rates}
\label{fig5-2}
\end{figure}

Analyzing the results in Table~\ref{tableII}
it is evident that integrating the transformer block can improve the activity detection accuracy of the target speaker.
Besides, the length of the segmentation window also affects the performance of TSVAD, which is optimal at 1.5s. 
When the window length is 3s, the fine granularity of generating target speaker activity labels is insufficient. When the window length is 0.75s, the speaker information of $E_Y$ is unreliable. 
The appropriate window length is one of the parameters to ensure the performance of speaker activity detection.

\subsection{Performance on different overlap ratios}

Our further evaluate the performance of each model across various overlap rates.
Fig.~\ref{fig5-1} illustrates that the Cascade approach 1 and pBSRNN demonstrate superior SI-SNR performance. 
In contrast, VBx and A-TSVAD exhibit a marked performance decline with increasing overlap, underscoring the heightened sensitivity of speaker diarization tasks to overlap variations.
Fig.~\ref{fig5-2} clearly depicts a significant decline in the performance of VBx with increasing overlap rates. The proposed Cascade approach 1 exhibits higher robustness. Their performance at an overlap rate of OV40 surpass that at OV30. This superior performance is attributed to their enhanced capability in managing overlapping audio. 

These results emphasise the effectiveness of the Cascade approach 1 in the C-TSE framework when dealing with different overlap rates.

\section{Conclusion}
In this paper, we introduce the C-TSE framework, a novel extension of traditional TSE designed to accommodate more intricate scenarios. These scenarios encompass multiple speaker environments, variable ratios of audio overlap, and situations where the target speaker may be absent.
In contrast to conventional clustering-based approach, the A-TSVAD model goes one step further.
We undertake a comprehensive investigation into the integration schemes of TSVAD and TSE, demonstrating that a sequential cascade fusion strategy further ameliorates extraction performance metrics.
This study has a limitation in that it focuses on matching enrollment and mixture record conditions.
We will investigate TSE under different recording conditions, such as recording equipment, room size, and speaker age.

\printbibliography

\end{document}